%% file: main.tex
\documentclass{article}

\usepackage{arxiv}
\usepackage{amsmath,amsthm,amssymb}
\usepackage{graphicx}
\usepackage{wrapfig}
\usepackage{lscape}
\usepackage[utf8]{inputenc} 
\usepackage[T1]{fontenc}    
\usepackage{hyperref}       
\usepackage{url}            
\usepackage{booktabs}       
\usepackage{amsfonts}       
\usepackage{nicefrac}       
\usepackage{microtype}      
\usepackage{lipsum}
\usepackage{wrapfig}
\usepackage{balance}
\usepackage{lscape}
\usepackage{amsfonts}
\usepackage{booktabs}
\usepackage{rotating}
\usepackage{siunitx}
\usepackage{threeparttable}
\usepackage{geometry}
\usepackage{longtable}
\usepackage{caption}

\DeclareUnicodeCharacter{2060}{\nolinebreak}

\title{Survey of ETA prediction methods in public transport networks}

\author{
Thilo ~Reich \\
Department of Computing and Informatics\\
Bournemouth University\\
Poole, United kingdom\\
\texttt{treich@bournemouth.ac.uk} \\
   \And
Marcin ~Budka \\
Department of Computing and Informatics\\
Bournemouth University\\
Poole, United kingdom\\
    \And
Derek ~Robbins \\
Faculty of Management,\\
Bournemouth University\\
Poole, United kingdom\\
    \And
David ~Hulbert \\
We  Are  Base  \\
Passenger  Technology Group\\
Bournemouth, United kingdom\\
}

\begin{document}
\maketitle

\begin{abstract}

The majority of public transport vehicles are fitted with Automatic Vehicle Location (AVL) systems generating a continuous stream of data. The availability of this data has led to a substantial body of literature addressing the development of algorithms to predict Estimated Times of Arrival (ETA). Here research literature reporting the development of ETA prediction systems specific to busses is reviewed to give an overview of the state of the art. Generally, reviews in this area categorise publications according to the type of algorithm used, which does not allow an objective comparison. Therefore this survey will categorise the reviewed publications according to the input data used to develop the algorithm. The review highlighted inconsistencies in reporting standards of the literature. The inconsistencies were found in the varying measurements of accuracy preventing any comparison and the frequent omission of a benchmark algorithm. Furthermore, some publications were lacking in overall quality. Due to these highlighted issues, any objective comparison of prediction accuracies is impossible. The bus ETA research field therefore requires a universal set of standards to ensure the quality of reported algorithms. This could be achieved by using benchmark datasets or algorithms and ensuring the publication of any code developed. 
\end{abstract}

\section{Introduction}
The UK has seen a constant rise in vehicles on its roads since personal vehicles have become available, which resulted in a 7 fold increase in traffic on British roads between 1950 and 2016~\cite{DepartmentofTransport2016Road2016}. This has naturally led to an increase in congestion felt by all road users. In a recent report, it was estimated that UK travellers spent 10\% of their driving time in gridlock~\cite{Cookson2017INRIXScorecard}. The reduction of congestion became a key priority as it will have a positive impact on the environment, the economy and will reduce commute times. This has been recognised for example in the UK government's `Road to Zero' strategy aiming to tackle emissions from road usage. The biggest environmental and societal impact can be achieved if the public is encouraged to use alternative modes of travel instead of private cars~\cite{Xia2015Traffic-relatedAustralia}. This review is focused on public buses as 4.44 billion bus journeys are made annually in the UK. Despite this, the patronage is declining and better Estimated Time of Arrival (ETA) predictions could play a role in slowing down this trend. It has been shown that even small changes can have a significant impact on the overall congestion of a city as highlighted by the fact that reducing daily commutes from specific neighbourhoods by only 1\% can cut delays for all road users by as much 18\%~\cite{Wang2012UnderstandingAreas}. Even if the cancelled commutes are randomly selected, delays can still be reduced by as much as 3\%. To encourage road users to change their mode of transportation, public transport has to be convenient and reliable. Punctuality and timeliness of the journey have the biggest impact on passenger satisfaction~\cite{TransportFocus2017BusReport}. Non-surprisingly, the most frequently requested improvements by passengers are accurate travel times both pre-trip and during the journey, especially for passengers using public transport to commute~\cite{Grotenhuis2007}⁠. To provide this punctuality buses should ideally adhere to a timetable, that has been carefully designed to allow the bus to meet it without introducing too much buffer times to lengthen the journey unnecessarily. However, this is often difficult and therefore it is crucial to accurately predict the arrival times of vehicles. This will improve passenger satisfaction even if the vehicle is late as passengers, in general, do not mind waiting within reason as long as they know for how long~\cite{Mishalani2006}. Furthermore, reliable real-time travel information delivered to passengers reduces the perceived waiting time for bus passengers as well as the actual waiting time as passengers can arrive more closely to the departure time~\cite{Watkins2011}. Furthermore, it will allow developing new smart applications allowing to offer personalised journey suggestions to the traveller. Because buses are affected by a large number of external influences such as weather, traffic conditions, passenger loads~\cite{Xinghao2013} and other types of disruptions, predicting their arrival is challenging and therefore currently not very accurate~\cite{Salvador2018}. Methods to predict ETA can include simple historical averages or be based on statistical models. Therefore, such techniques applied to bus ETA predictions can be expected to drastically improve the current performance. However, due to the complexity of the ETA prediction machine learning methods have become increasingly popular~\cite{Choudhary2016}. In recent years, Artificial Neural Networks (NN) have revolutionised a number of other domains. Therefore NNs should be expected to have the same potential when applied to bus ETA prediction problems. A comprehensive review specifically investigating NN applications in public transport~\cite{Pekel2017} found that only 16\% (12) addressed ETA of buses, whereas the rest of the studies applied the techniques to other modes of transport. This suggests that the area of bus ETA prediction using NNs might be underrepresented in the context of public transport research. This relative absence of NN to predict bus ETA is striking as NNs has revolutionised other areas of data science such as image and speech recognition. Nowadays the majority of buses have onboard Automatic Vehicle Location (AVL) systems, which are equipped with GPS sensors and transmit the location of the bus at frequent intervals, typically ranging between 20 and 60s. The availability of vehicle locations are the basis for any ETA prediction and are readily accessible through the AVL systems and do not necessarily need any additional investment in static sensors. The general approach of published reviews of ETA predictions methods is either to categorise by area of application or by the technique used as in~\cite{Pekel2017} or by the applied algorithm~\cite{Altinkaya2013,Choudhary2016}.
This review will asses the current literature concerning ETA prediction for buses. In doing so it will demonstrate a more informative categorisation than commonly used to review the literature and address shortcomings of the reporting standards. 

\section{Categorisation of ETA prediction algorithms}

\begin{wrapfigure}{r}{0.6\linewidth}
\fbox{\includegraphics[width=10cm, trim={2cm 1.5cm 1.5cm 1.5cm},clip]{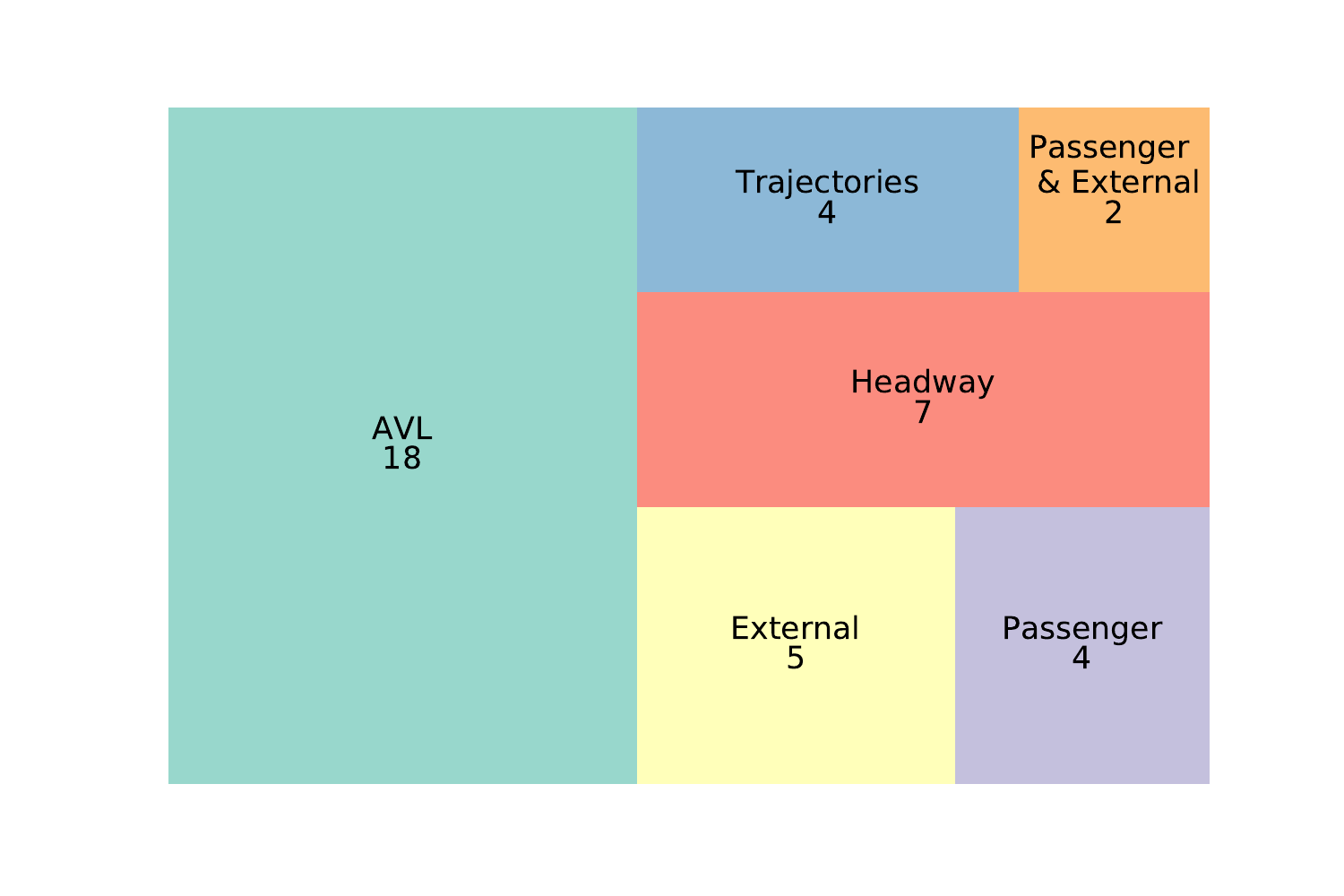}}
\caption{Treemap showing the proportions of the input features used in the reviewed publications.}
\label{fig:Feature}
\end{wrapfigure}

ETA prediction methods are commonly reported as categorised literature reviews based on the type of algorithm used as suggested in~\cite{Choudhary2016, Treethidtaphat2017}. This categorisation is not necessarily informative for the reader, as the algorithms can be developed based on different background information -- different input features such as locations, speed and passenger load of the vehicle are used to develop the algorithm, which prevents any meaningful comparison. Therefore, approaches that were developed using only AVL data should in most cases not be compared to methods accounting additionally for passenger load as well as weather conditions, even if it might be based on the same algorithm. Typically AVL data includes vehicle positions and schedule and route identifiers but can include more information depending on the provider. This would compare algorithms relying on entirely different extent of information thus preventing a meaningful interpretation. As this review's focus lies on the prediction of bus ETA, the

\begin{wrapfigure}{r}{0.6\linewidth}
\fbox{\includegraphics[width=\linewidth,clip]{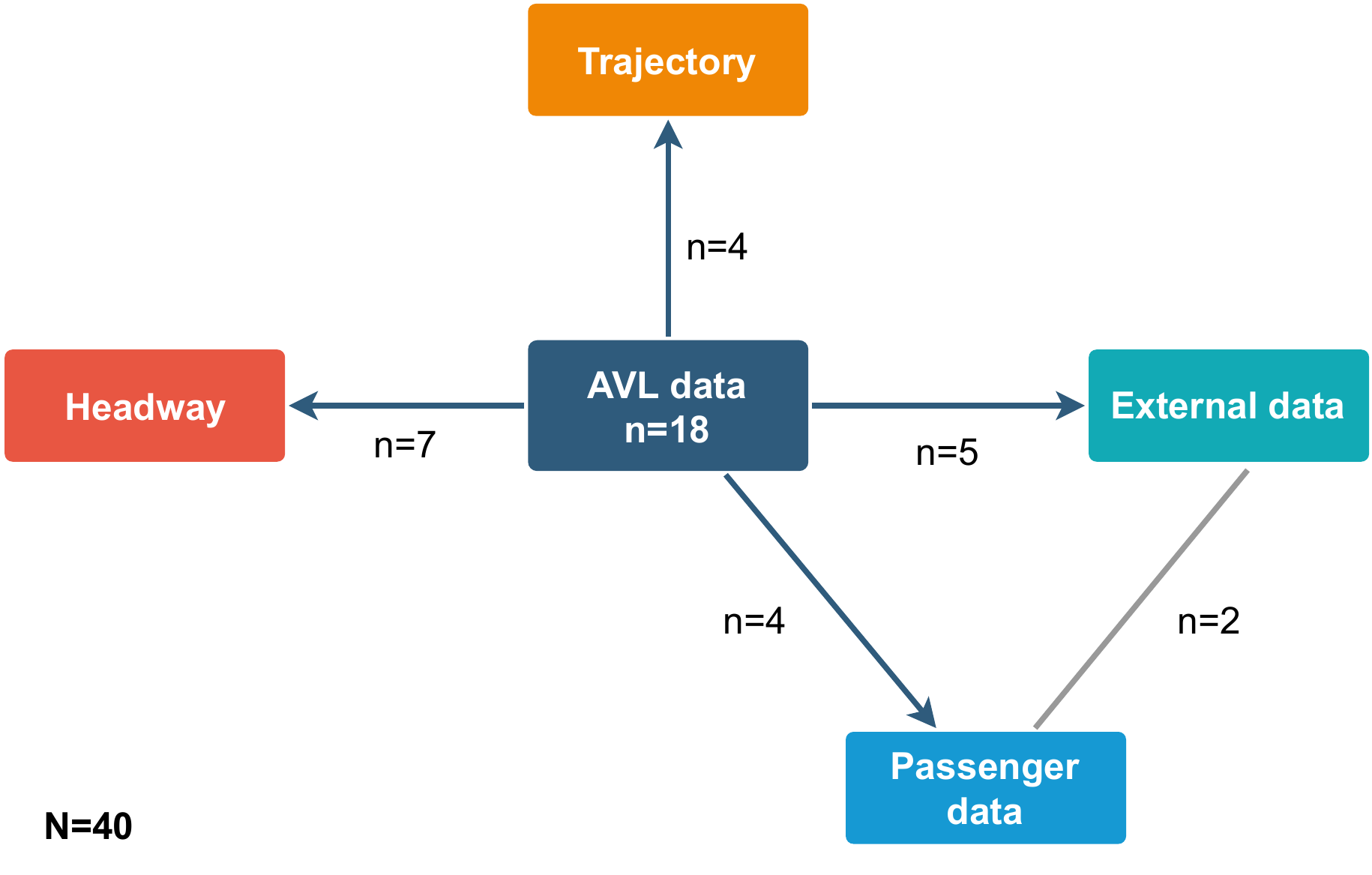}}
\caption{Categories used to review the literature based on feature types.}
\label{fig:Types}
\end{wrapfigure}

reviewed studies are categorised based on the nature of input features used. The most basic requirement of input features to predict ETA are sequences of timestamped GPS coordinates recorded by AVL systems (n=18, Figure~\ref{fig:Feature}). These features were used by all 40 reviewed publications (also see supplementary 1). The different feature sources were found to be from external data such as information about the traffic or the weather (n=5), Passenger information such as load and embarking and disembarking numbers (n=4), and a combination of all three aforementioned sources (n=2). A separate group of studies used AVL information from the bus to be predicted in combination with AVL data of other buses serving the same route to calculate the Headway (n=7).

\subsection{AVL as sole data source}

A minimum requirement to allow any ETA prediction is the knowledge about the position of a vehicle, hence most reviewed studies used AVL data from onboard devices. The only exception was~\cite{Zhang2015}, where the locations were recorded using a modified mobile phone as the buses were not equipped with a GPS system. The reviewed studies used data, which included time-stamped positions of the buses and in some cases, additional information was explicitly calculated such as average speeds or dwell times. Therefore, this central group of features was the most common and thus also includes the widest range of applied techniques. The simplest ETA prediction based solely on AVL data are historical methods using the average speed from historical records to predict the arrival time at a destination~\cite{Lin1999}. Naturally these cannot account for any fluctuations and thus perform with up to 9.3\% lower accuracy compared to more intricate methods such as Kalman Filters (KF) \cite{Kumar2017}. Attempts to improve simple historical mean based algorithms, such as accounting for timed stops at which the timetable has deliberate waiting times, reduce the prediction deviation by 0.8~\cite{Lin1999}. Another approach was used in which the prediction was made using the historical average updated with exponential smoothing for several short sections of the route, which are then combined to give the total travel time~\cite{Meng2017}. 
In the search for an algorithm with better performance and lowest computational impact~\cite{Maiti2014}, compared a historical average method, Artificial Neural Networks (NN), and Support Vector Machines (SVM). The results suggest that the NN did outperform historical methods with a minuscule advantage although the exact value of the improvement is not reported. The author's conclusion is that as the NN and the historical method perform similarily, yet the NN requires more intensive training and longer prediction times, the historical method is superior~\cite{Maiti2014}. However, the overall consensus of the literature regarding historical methods is that their performance is low~\cite{Shalaby2003,Shalaby2004,Vanajakshi2009TravelBuses}.

Kalman Filter (KF) is a statistical method that has been applied to bus arrival times ~\cite{Dailey2001TransitImplementation,Cathey2003, Padmanaban2009} and was found to perform with better accuracy in comparison to historical methods (maximum relative error of 0.543 of the historical approach and 0.087 for the Kalman Filter)~\cite{Shalaby2003,Shalaby2004,Vanajakshi2009TravelBuses}. Autoregressive Integrated Moving Average (ARIMA) exploit the information contained in the timeseries and was used in one example with acceptable results compared to the ground truth (MAPE=3.88-6.42\% depending on direction). Unfortunately, it was not compared to any other methods making it difficult to objectively bring this method into context~\cite{Napiah2009ArimaPrediction}. A direct comparison of historical methods to Linear Regression (LR) in~\cite{Shalaby2003,Shalaby2004} showed that LR performed with up to 6.7 times lower error than historical methods. However, KF performed up to 3.95 times better than LR. This study is the only example of a direct comparison of KF and LR. When compared to regression models, NNs generally perform with higher accuracy when trained on the same dataset~\cite{Amita2015}. Historical-based and regression methods do not cope well with fluctuations~\cite{Treethidtaphat2017} and variations of travel times are highly likely at peak times in the urban environment. Therefore, non-linear methods such as NN should intuitively perform better when used with more complex data with higher variation. Pan et al.~\cite{Pan2012} used an NN to predict the average speed for the remaining distance to the destination, improving the accuracy compared to a historical algorithm by 5.7\%. Similarly, in Houston a NN outperformed historical and regression models~\cite{Jeong2004a}. Interestingly, this study also found that the improvement although drastic compared to the historical algorithm was less pronounced in the suburban areas presumably due to congestion. This also materialises from the findings by~\cite{Julio2016}, that overall NNs performed significantly better. An exception was heavy congestion where historical approaches were more accurate than NNs. Further investigations found that the NN overestimated speeds in slow conditions and underestimated travel times at high speeds. Surprisingly, the information whether a bus was currently on a bus lane did not influence this behaviour. Generally, ETA predictions are made by estimating the absolute number of minutes until arrival or the travel speed. In a unique approach, \cite{Kee2017} treated the estimation as a classification problem by predicting the 1/4 h when the bus will arrive. In their experiments, an NN based approach performed 8\% better than Decision Trees, Random Forests (RF) and Naive Bayes. The ensemble approach was also used to combine several NNs where the parameters such as the number of layers and neurons ware generated randomly and the best performing was included into one ensemble~\cite{Chen2018}. Unfortunately, the authors do not report the exact architecture of the final NNs. As the number of layers could have ranged between 1-5 this could be an example of a deep neural network if this information was known.  

\input{table1}

The relative absence of deep learning approaches is striking in the context of bus-ETA prediction. A reason could be the reported behaviour that NNs with a single hidden layer outperformed NNs with two or three layers thus suggesting that shallow NNs might be sufficient or even desirable to predict bus ETAs~\cite{Heghedus2017}. However, as ETA prediction is a sequential problem it can be expected that Recurrent Neural Networks (RNN) and their derivatives will perform better. The reason for this is the design specifically tailored to sequential data, where the depth of the network is linked to the length of the sequence~\cite{Lipton2015ALearning}. A similar conclusion was made by~\cite{Chen2004a} who found in a comparison of NN architectures that the more hidden layers a network had the less likely it was to generalise. In contrast~\cite{Treethidtaphat2017} used a NN with 4 hidden layers reporting excellent performance compared to ordinary least square regression. As this study does not report on any NNs with different depths the results are difficult to interpret. Generally arrival times are predicted for designated bus stops, however in some public transport systems buses can be flagged down anywhere on their route. In a study in Bangkok a 4 layer deep neural network was used to improve the prediction of arrival in comparison to a regression model resulting in an error reduction of 55\%~\cite{Treethidtaphat2017}. The dilemma of choosing a suitable NN architecture has led~\cite{Khosravi2011AIntervals} to use a genetic algorithm (GA) to select the best performing architecture. As it is unlikely that any model will be able to perform with the same accuracy under every condition, some authors have tried to overcome this limitation by using hybrid methods. Such an example is a combination of a SVM and KF by~\cite{Yu2010HybridStation}, where the SVM predicted baseline values used for the KF prediction. The SVM-KF hybrid achieved 11.1\% higher accuracy than a NN-KF hybrid. Nevertheless, the most commonly used method in the context of ETA prediction are NNs. Considering that the majority of publications are using shallow networks and there are few examples of deep learning architectures (3/19) this poses the question whether the reason is that these architectures do not work in this context. A possible reason could be that the studies focused on input features of one bus route thus limiting the data complexity compared to an approach, using the network-wide state of all buses as input.

\subsection{Trajectory based methods}

\begin{wrapfigure}{r}{0.6\linewidth}
\fbox{\includegraphics[width=\linewidth,clip]{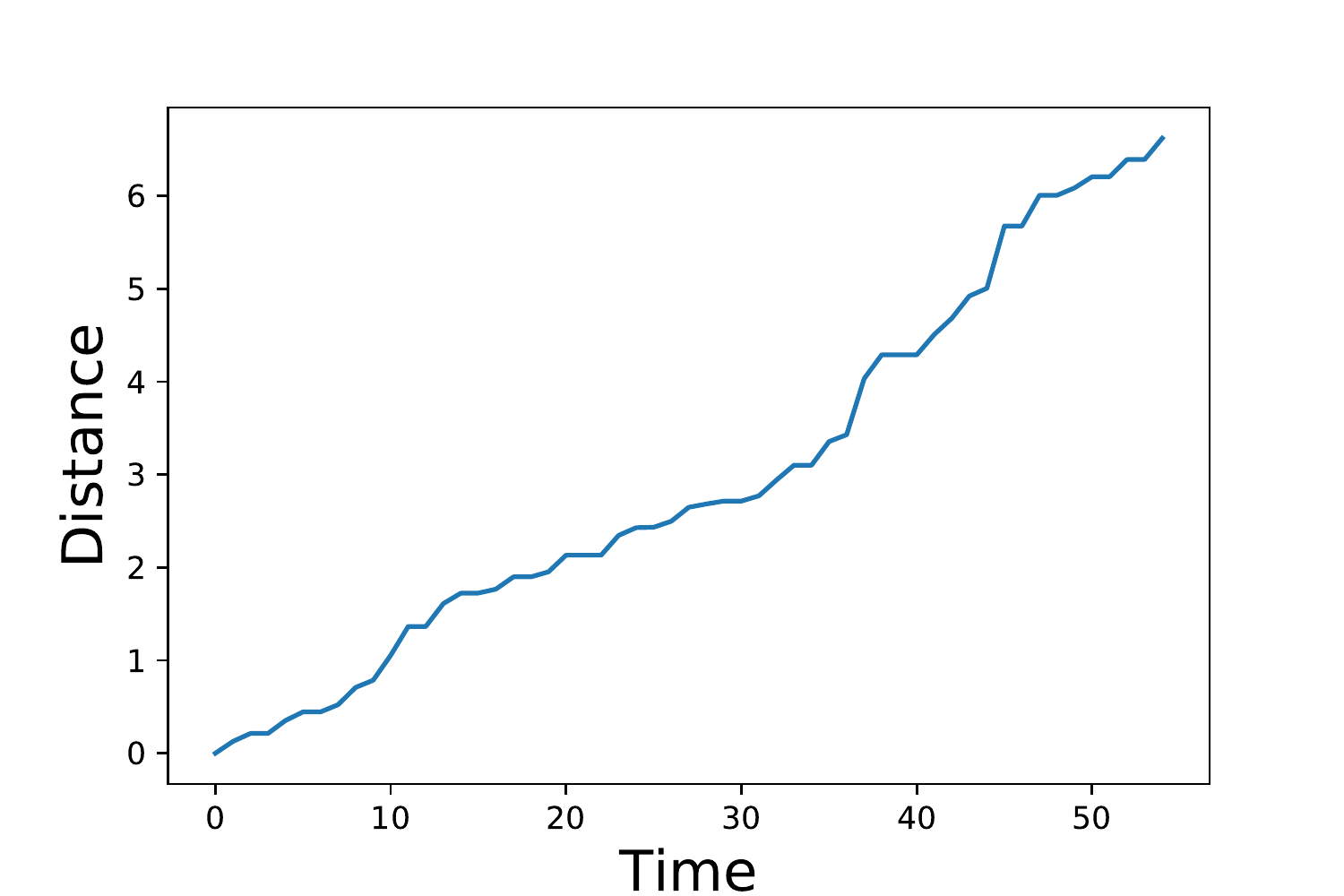}}
\caption{Example of a bus trajectory illustrating the travelled distance over time.}
\label{fig:Trajectory}
\end{wrapfigure}

Trajectory based methods use historical trajectories of a bus line i.e. the distance travelled by a bus over time (see Figure~\ref{fig:Trajectory} for an example). The estimate is being made by comparing the current trajectory of a bus with those of the past and using the most similar trajectory as a prediction. The choice of an appropriate trajectory is made by different algorithms. 

One such example is the work described by~\cite{Dong2013}, who select the most similar trajectory using a k-Nearest Neighbour (kNN) algorithm. In this study, it was found that the kNN algorithm outperformed a NN approach for long term prediction. Interestingly, this approach did not perform well on short distances below 3 km and the authors reverted to use the average speed of all buses travelling on the same road segment as a prediction. Similarly, \cite{Kumar2017b} used a kNN classifier to select historical trajectories which were then fed into a KF to predict the bus travel time. In a modification~\cite{Xu2017} grouped the trajectories into categories based on road segments and time of day. Then the prediction was made by comparing the progress of the current bus to the historical trajectories corresponding to the time and section the bus is presently travelling on. This approach was used in order to reduce computational cost and was shown to outperform SVM and NN trajectory matching approaches. In a comparison of different methods applied to trajectories, kernel regression was superior to both LR and kNN methods~\cite{Sinn2012}.

\subsection{AVL and headway information}

As the progress of a bus is naturally dependent on traffic flow, information about the state of the forward traffic should improve the accuracy of any algorithm. As all the reviewed methods used AVL data this allows using these data from previous buses as an indication of the traffic ahead. The distance or time to the preceding buses is called headway and was used in 6 out of the 40 reviewed studies. An example specifically looking at bus stops served by multiple routes, showed that the best accuracy could be achieved if not only the weighted headway to preceding buses of the same route but also those to buses of other lines were included. This was true when the prediction was made using an SVM, interestingly, excluding the running time of the same line resulted in the best prediction using a NN but was still outperformed by the SVM~\cite{Yu2011}. 

In contrast, when accounting for travel times of the preceding buses on a virtual road, an NN solution was found to perform better than an SVM~\cite{Hua2017}. However, \cite{Hua2017} used different features as well as 2 hidden layers instead of 1 thus making a comparison difficult. A further study found that an SVM had slightly better accuracy than NN models and KFs. The error was nearly halved if KF was used upstream of either model to account for dynamic changes. Also, in this case, the SVM-KF model was slightly superior to the NN-KF approach~\cite{Bai2015}. 

\cite{Yin2017} found that overall both NNs and SVMs resulted in a prediction error around 10\% although with minimal variations over the course of the day and in different city environments. A genetic algorithm was used to determine the best architecture for a NN, which resulted in an NN with 1 hidden layer and 5 hidden units (3-5-1). This is the same structure as~\cite{Yu2011} whereas~\cite{Bai2015} used 6 hidden units. The described works are very consistent in the selection of network depth as well as their findings. 

An advancement from a simple NN was presented by~\cite{Lin2013Real-TimeChina}, who used a hierarchical NN. This approach trained sub-NNs for clusters based on the day of the data collection as well as the delay level at the time of collection. These were then combined into a hierarchical NN which performed better than the conventional NN and KF. Other hierarchical methods are Random Forests which surpassed SVM, kNN and LR. The error was further reduced by 1.3\% if the RF was trained on datasets preselected using a kNN approach accounting for the intuition that under similar circumstances the travel time will be similar~\cite{Yu2017}. The methods described in this section use headways as additional inputs to AVL data. However, one method instead used queuing theory. The so-called snapshot method simply uses the travel time of the last bus traversing the same segment as a prediction. To minimise the effect of outliers on this approach, different RF based methods were used to get the final prediction based on the snapshot design~\cite{Gal2017}. 

\subsection{AVL and external data}

As any road user knows, progress in traffic depends on many external influences, such as weather or traffic volume. This is also true for buses and has been addressed in a number of studies. The weather conditions have been taken into account in two studies. 
One basic example including weather influences used a SVM to make ETA predictions based on data from the last 30 days. These predictions are then stored and used as predictions for all journeys of the next day. Naturally, this will not account for any sudden changes in external conditions. Regrettably, this study does not compare the method to any other approaches thus making it impossible to objectively evaluate it~\cite{Li2018}. Similarly, ~\cite{Junyou2018ApplicationPrediction} used an SVM to predict ETAs based on the last four days in order to predict the 5th. An interesting approach used cameras on overhead bridges to not only count bus traffic but also the speed of taxis as these can use the same routes as buses and unsurprisingly found that their speed is the same in heavy traffic. Furthermore, it was found that the prediction solely based on the information from the static cameras identifying the bus was more accurate than if it was using only GPS recordings. The authors did not combine both in order to investigate whether this would improve the overall performance although this would have been an insightful addition to their research~\cite{Xinghao2013}. Again, these methods were not compared to any alternative approaches. A combination of both weather and traffic state was used in a hybrid method. The reasoning is that NNs are often poor at accounting for disruptions, therefore, a system was used, employing an NN for traffic situations that appear to be `normal' in the sense that the system has encountered similar conditions before. If it appears to be an unseen condition the prediction is made using a KF. This improves the performance compared to an NN that is used for all conditions by 0.2 min error for the entire route (37min)~\cite{Zaki2013}. This highlights the crux that it is unlikely that one method will always perform best and it can be anticipated that different conditions will affect a model's performance. 

A preliminary report~\cite{Heghedus2017} describes attempts to use LSTMs to predict bus ETAs and including both traffic and weather data, but full results have not yet been published.

\subsection{AVL and Passenger data}

As public transport's purpose is to convey passengers, the customers themselves affect the progress of any bus. The number of passengers boarding will have an influence on the dwell time as well as on the frequency of stops made by the vehicle.

An interesting sensitivity analysis~\cite{Chen2004a} showed that the impact of dwell time on the ETA of a bus has an effect of 45\% whereas the day of the week played a 25\% role. In practice, it is difficult to include the exact number of passengers as this information is generally not collected automatically since tickets do not necessarily have information about the destination and passengers do not have to swipe for example a smart card when disembarking. However, if this data could be made available it should give information about future dwell times as more passengers require longer to disembark. 

Therefore, passenger numbers boarding and disembarking were included in an NN model that performed significantly better than LR with the same inputs~\cite{Wang2014}. Due to the difficulty of assessing the number of passengers an imaginative way used the microphone of a mobile phone installed on the bus to count the sound made when a smart card was swiped at the terminal by a passenger. This information was used to record the number of boarding passengers without any information about the number disembarking~\cite{Zhang2015}. In a comparison, ~\cite{Shalaby2003} found that a KF performed better if data including location and passenger load were included. This outperformed a time-lagged NN, as well as LR and a historical model. The same study was republished~\cite{Shalaby2004}. This model was later replicated and found to perform with the lowest accuracy when compared to NNs and Hierarchical NNs~\cite{Lin2013Real-TimeChina}. This illustrates the replication problem found in the current literature inhibiting any objective comparison of the proposed methods. 

\subsection{AVL and passenger and external data}

To account for as many external influences as possible several studies combined both data from external sources such as weather and traffic and information about the passengers. 

A NN-KF hybrid where the NN feeds into the KF was developed using features including weather (and more specifically precipitation), passenger loads, boarding and disembarking as well as AVL information. The hybrid did perform better than a conventional NN~\cite{Chen2004AData}. Generally, two methods of segmentation of a route exist: (1)~the stop based segmentation where the travel time between two stops is predicted, and (2)~the link based prediction where the travel time of a link consisting of several stop to stop segments is estimated. The travel time can either be predicted using a stop-based approach where the time needed from one stop to the next is predicted or a link-based method where the route between two stops is split into several shorter links and each link is predicted separately. In a comparison of the stop based and link-based ETA predictions using AVL data and traffic flow data as features, it was found that the stop based method performed with up to 2.7 times smaller error~\cite{Chien2002}.

section{Discussion}

The feature-based categorisation used in this review, allowed a better understanding of the applied methods to predict bus ETAs. The analysis highlighted several flaws in the current research that make the interpretation of the results challenging. A reliable comparison of the methods was not possible because the measures used to report the algorithm performance were inconsistent. Furthermore, one of the reviewed papers presented an algorithm without any comparison to other methods, thus preventing any objective assessment. Lastly, the reporting quality of some papers was inadequate. 
Following each point will be discussed individually. 

\subsection{Comparability}
As the accuracy and performance of any prediction model is of crucial importance, this has to be reported in a way that allows to replicate and compare the results. However, this is not possible in all cases as some authors report relative errors and no consistency in the reported parameters can be distinguished. The precondition that any developed machine learning algorithm should fulfil is verifiability and has been highlighted by a report of the Royal Society as one of the central importance~\cite{TheRoyalSociety2017MachineExample}. This has also been recognised in the healthcare sector where guidelines for the development and reporting of predictive models exist~\cite{Luo2016GuidelinesView.}. The difference in standards might be explained because ETA predictions do not affect the health or safety of a passenger and a spurious algorithm might at most cause inconvenience rather than physical harm. However, for an operating company, this might cause a loss of revenue because patronage might decline. Furthermore, the society as a whole might be subjected to more congestion, that could simply be reduced by providing accurate ETA predictions. Furthermore, the doctrine of science is replicability. The reproducibility crisis is most prominently known from psychological research~\cite{Baker2015OverTest} however due to its notoriety it is actively being addressed~\cite{Yarkoni2017ChoosingLearning}. It has also been identified as a problem in `harder' sciences such as biomedicine~\cite{Baker20161500Reproducibility} and also artificial intelligence~\cite{Hutson2018ArtificialVerify}. Although results gained from machine learning techniques might be considered to be hard evidence, because the final model is based on mathematical concepts, they suffer from similar problems as seen in psychology where the research is often subjective to the researcher. The similarities between the two fields are that the findings cannot usually be explained due to the `black box' effect. The field of psychology has now started to apply lessons from problems seen in machine learning research~\cite{Yarkoni2017ChoosingLearning}. A suggested way of addressing such problems is meta-science that could shed light on the true accuracy of findings~\cite{Schooler2014MetascienceCrisis}. However, this relies on comparable measurements of accuracy, which was not found in a large proportion of the reviewed literature. Therefore, comprehensive standards of reporting are urgently needed in the field of predictive bus transportation research.

\subsection{Comparison}
Leading on from the reproducibility problems is the lack of comparison to other methods found in a large proportion of studies (n=11, 27.5\%). This would not be a major issue if the same prediction measurements were described, however, as this is not the case such reports only allow limited comparison between the studies. The findings, therefore cannot be compared to other researcher's work and therefore can only be considered standalone reports of a method applied to a certain problem. Such studies do not even give information about any possible relative improvements to other currently employed methods. If the researchers had directly compared their approach to a preexisting or commonly used algorithm, the value of the findings would increase. The comparison to other methods is the only way of establishing a benchmark to which any improvement can be compared to. 
https://www.overleaf.com/project/5ca487c0504f2453fce07a0d
\subsection{Quality}
The third issue is related to the reporting standards and a few studies did not make it clear what architectures were used in the final algorithm or left leeway in the interpretation of their findings, by not explaining graphs or figures or because of discrepancies between values in the description compared to the presented figures.

\subsection{Conclusion}

\begin{wrapfigure}{r}{0.6\linewidth}
\fbox{\includegraphics[width=\linewidth, trim={2cm 1.5cm 1.5cm 1.5cm},clip]{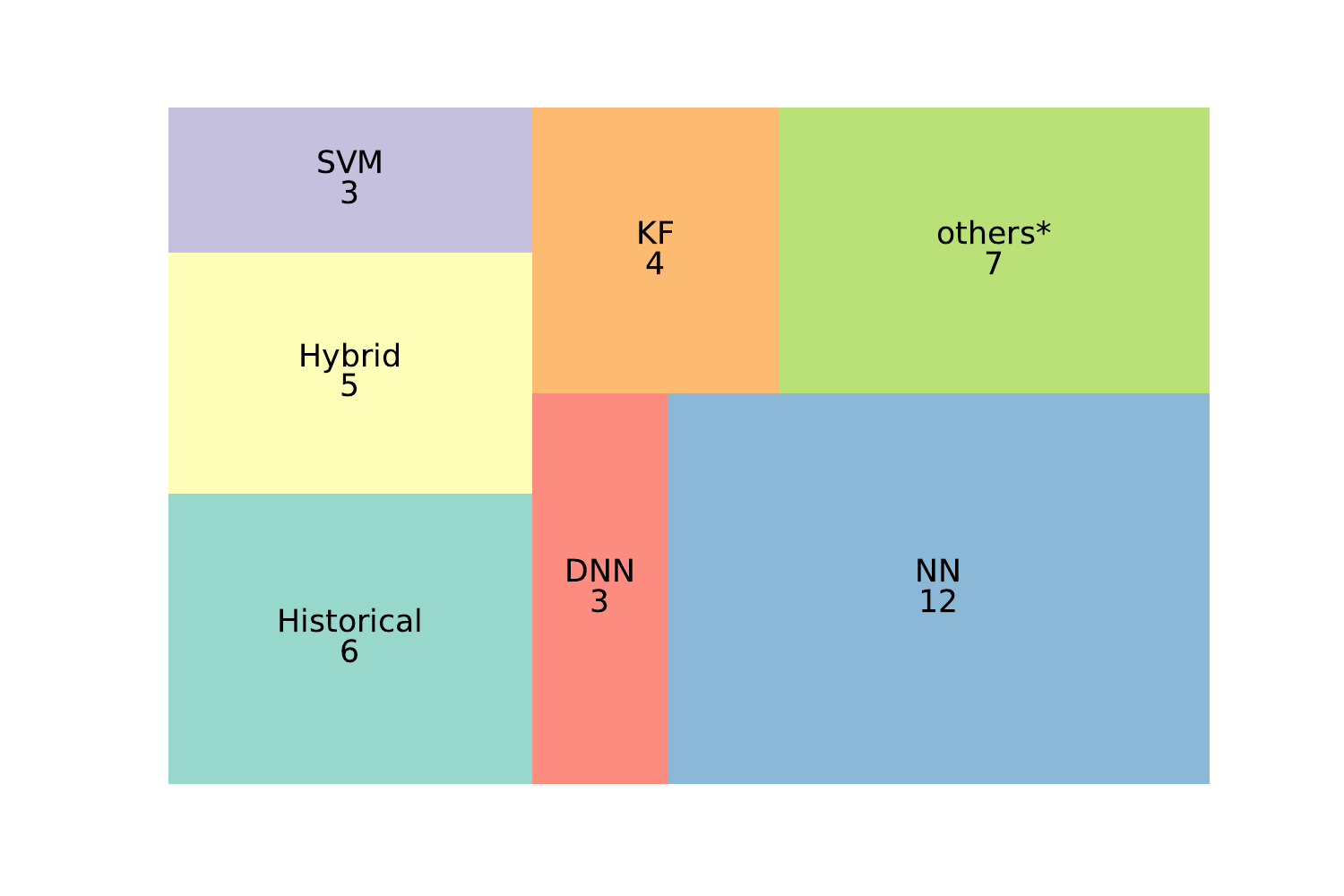}}
\begin{small}
* methods n \textless 2 such as RF, Bayesian Networks etc.
\end{small}
\caption{Proportion of each method used in the reviewed studies.}
\label{fig:Methods}
\end{wrapfigure}

This review highlighted some shortfalls of the current literature addressing the ETA prediction of buses. Overall NNs predominated (n=12, 30\%) the methods (Figure \ref{fig:Methods}). Also, deep learning approaches with more than 2 hidden layers have been used in 4 publications. However, in one approach an iterative selection of layer numbers and units was applied but the final layer number was not reported. 

It was telling that several studies found different algorithms performing better in different settings suggesting that there will not be one superior algorithm for all cases. Unfortunately, due to the highlighted shortcomings, it is not possible to identify the `best' method for each of the categories. Considering the popularity of NNs it appears to be the most widely used method suggestive of being the best performing and/or most universal. 

Interestingly, deep learning approaches are underrepresented and in some cases, it was found that 2 layer networks were performing better than deeper architectures. This could be due to the fact that NNs allow representing any nonlinear relationship between variables, in data with lower complexity. In general, the input features used consisted of data regarding one bus line and several variables directly linked to this line such as other vehicles travelling on the same route. It would be expected that deep learning approaches will be more successful in generalising more complex datasets for example if the entire network state is considered, including information about all vehicles.

Concluding it can be said that research into bus ETAs lacks consistency and uniform standards. Ideally, an approach similar to image classification or other areas could be used where a standard reference dataset is made available and used as benchmark performance test. Alternatively, if the used data was published alongside the used code this would help increase the comparability. Furthermore, it became clear that an industry-wide standard for reporting prediction accuracy is urgently needed.

\newpage

\bibliographystyle{unsrt}  
\bibliography{references}

\newpage

\newgeometry{left=2cm, bottom=1.5cm}
\begin{landscape}
\pagestyle{empty}

\nobalance
\subsection{Supplementary 1}

\input{table2}

\end{landscape}
 
\end{document}

%% file: table1.tex
\begin{wraptable}{r}{10cm}
\begin{small}

\centering

\begin{tabular}{SSSSSSSS} \toprule

    \textbf{Publication} & \rotatebox{90}{\textbf{AVL}}& \rotatebox{90}{\textbf{External}} & \rotatebox{90}{\textbf{Trajectories}} & \rotatebox{90}{\textbf{Passenger}} & \rotatebox{90}{\textbf{Headway}} \\

    \\ \midrule
    \textbf{Amita et al. (2015)}  \cite{Amita2015}  & $\bullet$  \\
    \textbf{Bai et al.  (2015)} \cite{Bai2015} & $\bullet$ & & & & $\bullet$\\
    \textbf{Chen (2004a)} \cite{Chen2004a} & $\bullet$ \\
    \textbf{Chen et al. (2004b)} \cite{Chen2004AData} & $\bullet$ & $\bullet$ & & $\bullet$\\
    \textbf{Chen  (2018)} \cite{Chen2018} & $\bullet$\\
    \textbf{Chien et al. (2002)} \cite{Chien2002} & $\bullet$ & $\bullet$ & & $\bullet$\\
    \textbf{Dailey et al. (2001)} \cite{Dailey2001TransitImplementation}   & $\bullet$\\ 
    \textbf{Deng et al. (2013)} \cite{Deng2013} & $\bullet$ & $\bullet$\\
    \textbf{Dong et al. (2013)} \cite{Dong2013}  & $\bullet$ & & $\bullet$ \\
    \textbf{Gal (2017)} \cite{Gal2017}  & $\bullet$ & & & & $\bullet$ \\
    \textbf{Heghedus (2017)} \cite{Heghedus2017} & $\bullet$ \\ 
    \textbf{Hua et al. (2017)} \cite{Hua2017}  & $\bullet$ & & & & $\bullet$ \\
    \textbf{Jeong \& Rilett (2004)} \cite{Jeong2004a} & $\bullet$ \\
    \textbf{Julio et al. (2016)} \cite{Julio2016} & $\bullet$ \\
    \textbf{Junyou et al. (2018)} \cite{Junyou2018ApplicationPrediction} & $\bullet$ & $\bullet$\\
    \textbf{Kee et al. (2017)}  \cite{Kee2017} & $\bullet$ \\ 
    \textbf{Khosharavi et al. (2011)} \cite{Khosravi2011AIntervals}  & $\bullet$ & \\
    \textbf{Kumar et al. (2017)} \cite{Kumar2017}  & $\bullet$ & & $\bullet$ \\
    \textbf{Li (2018)} \cite{Li2018}  & $\bullet$ & $\bullet$\\
    \textbf{Lin \& Zeng (1999)}  \cite{Lin1999} & $\bullet$ \\
    \textbf{Lin et al. (2013)}  \cite{Lin2013Real-TimeChina} & $\bullet$ & & & & $\bullet$ \\
    \textbf{Maiti et al. (2014)}  \cite{Maiti2014} & $\bullet$ \\
    \textbf{Meng et al. (2017)} \cite{Meng2017}  & $\bullet$ \\
    \textbf{Nappiah et al. (2009)} \cite{Napiah2009ArimaPrediction}  & $\bullet$ \\
    \textbf{Padmanaban et al. (2009)} \cite{Padmanaban2009} & $\bullet$ \\
    \textbf{Pan et al. (2012)} \cite{Pan2012} & $\bullet$ \\
    \textbf{Shalaby \& Farhan (2003)} \cite{Shalaby2003} & $\bullet$ & & & $\bullet$\\
    \textbf{Shalaby \& Farhan (2004)} \cite{Shalaby2004} & $\bullet$ & & & $\bullet$\\
    \textbf{Sinn et al. (2012)}  \cite{Sinn2012} & $\bullet$ & & $\bullet$  \\
    \textbf{Treethidtaphat et al. (2017) } \cite{Treethidtaphat2017} & $\bullet$ \\
    \textbf{Vanajakshi et al. (2009)}  \cite{Vanajakshi2009TravelBuses} & $\bullet$ \\
    \textbf{Wang et al. (2014)}  \cite{Wang2014} & $\bullet$ & & & $\bullet$\\
    \textbf{Xinghao et al. (2013)} \cite{Xinghao2013} & $\bullet$ & $\bullet$\\
    \textbf{Xu (2017)} \cite{Xu2017}  & $\bullet$ & & $\bullet$  \\
    \textbf{Yin et al. (2017)} \cite{Yin2017}  & $\bullet$ & & & & $\bullet$\\
    \textbf{Yu et al. (2010)} \cite{Yu2010HybridStation} & $\bullet$ \\
    \textbf{Yu et al. (2011)} \cite{Yu2011}  & $\bullet$ & & & &  $\bullet$ \\
    \textbf{Yu et al. (2017)} \cite{Yu2017} & $\bullet$ & & & & $\bullet$ \\
    \textbf{Zaki et al (2013)} \cite{Zaki2013} & $\bullet$ & $\bullet$\\
    \textbf{Zhang et al. (2015)}  \cite{Zhang2015} & \multicolumn{1}{r}{$\bullet$\textsuperscript{1}}  & & & $\bullet$ \\ \bottomrule

\end{tabular}

\begin{small}
\begin{flushleft}
\noindent \textsuperscript{1} authors used a modified smartphone instead of a commercial AVL system.
\end{flushleft}
\end{small}

  \caption{\label{tab:Features}\small The input features used by each publication indicated as points.}



\end{small}
\end{wraptable} 

%% file: table2.tex
Download

    Source

    PDF

Actions

       Copy Project
       Word Count

Sync

       Dropbox

       Git

       GitHub

Settings
Compiler
Main document
Spell check
Auto-complete
Auto-close Brackets
Code check
Editor theme
Overall theme
Keybindings
Font Size
Font Family
Line Height
PDF Viewer
Hotkeys

       Show Hotkeys

ETA Literature Review ArXiv
T

\begin{centering}
 \begin{small}
\begin{longtable}{ p{2cm} p{2cm} p{4cm} p{1.5cm} p{3cm} p{4cm} p{2cm}} 

\toprule

    \textbf{Publication} & \textbf{Type} & \textbf{Architecture}& \textbf{Evaluation} & \textbf{Accuracy} & \textbf{Features} & \textbf{Comparison} \\

    \\ \midrule
    \endhead

    \textbf{ Amita et al. (2015) \cite{Amita2015}}  &  NN &   NN 3-5-1 and 3-15-1, output = travel time & 
        RMSE, MAPE, R2 &   NN: MAPE = 6.527\%, LR: MAPE = 16.234\% &    dwell time, delays, distance between stops & NN \textless LR \\ \midrule
    
    \textbf{Bai et al.  (2015) \cite{Bai2015}}   & SVM / NN adjusted with KF &     NN-KF: 8-6-1, output= travel time &   
    MAE, MAPE, RMSE &   
    MAPE for best road segment: NN and SVM: 10\%, NN-KF and SVM-KF: ~4\%,  KF: 10.68
    &  Time, Road segment, weighted average of travel time of other buses of other lines, travel time of preceding bus.   & NN-KF / SVM-KF  \textless NN / SVM  \textless KF\\ \midrule

 \textbf{ Chen et al. (2004a) \cite{Chen2004a}}   & NN &  
 
 Compared activation functions: hyperbolic tangent, tanh linear, sigmoid, sigmoid linear, sigmoid sigmoid: NN1; hidden layers= 1-2, number of neurons are not reported  & Average error &   prediction varies within 15\% of travel time. &  Cumulative dwell time, day of week, trip pattern.  & \textless comparison of different activation function combinations \\ \midrule    
    
 \textbf{ Chen et al. (2004b) \cite{Chen2004AData}}   & NN 
 
 & NN; 18-[4-6 neurons]-1 , output travel time between two points. NN is dynamically adjusted by KF incorporating the latest current arrival times for unique schedule patterns .   &  
 MSE, RMSE &   
MSE for best pattern; NN-KF: 0.009, NN: 0.016.  
 &  precipitation, time of day door opening/closing, stop sequence Ids, trip status, Coordinates, dwell time, stop distance number of passengers boarding disembarking, travel time between two consecutive stops, arrive passenger load and leave passenger load &  
 NN with dynamic adjustment \textless  NN \\ \midrule

 \textbf{ Chen  (2018) \cite{Chen2018}}   & NN potentially DNN &     
   Generated 10 NNs with random number of hidden layers between 1-5, and random number of neurons up to 7. Trained separate NN for urban and rural traffic and used 9 NN ensemble. & Average accuracy &   
    Average Accuracy: NN ensemble (architecture not reported): 94.75\%, NN: 94.65\%, LR: 94.42\%, Statistical mean: 94.08\% or LR. &  Historical stop to stop travel time. & NN ensemble \textless NN \textless LR \textless Statistical mean \\ \midrule    

 \textbf{ Chien et al (2002) \cite{Chien2002}}   & NN & NN link based: 4-6-1,  NN stop based: 6-7-1 
 & RMSE, sum of squares errors (SSE) & SE for; link based NN: 0.0965, stop based: 0.041 &  
 Link based NN: distance on link, traffic volume, link speed, link delay, link queue, passenger demand. 
 Stop based NN: distance between stops, mean traffic volume, std of volume, mean link speed, std of link speed, mean link delay, std link delay intersections, demand. & NN stop and link \textless NN stop base \textless NN link based \\ \midrule
    
 \textbf{ Dailey et al. (2001)} \cite{Dailey2001TransitImplementation}   & KF &      &   
    Kologormonov - smirnov &    &  Locations  & No comparison\\ \midrule

 \textbf{ Deng \& He (2013) \cite{Deng2013}}   & Bayesian network & Used traffic state (average speed for a section) as parent and arrival as child node. &   
    MAPE, MAE, RMSE &   MAPE=0.195, MAE=39.09, RMSE=49.14 & Road state & No comparison \\ \midrule
    
 \textbf{ Dong et al. (2013) \cite{Dong2013} }   & kNN &    
 Average speed of all buses that passed a point over the last 10 mins for predictions below 3km. NN: 18-37-15 where the output applies to all bus stops for distances \textgreater 3km.  
 
 &   APE (Long distance), AE (short distance) &   Better performance on long distances of kNN than NN no values reported. KNN: APE  \textless 12 \% mean 7\%. &  Trajectories &  kNN \textless NN \\ \midrule
    
 \textbf{ Gal et al. (2017 \cite{Gal2017}) }   & Snapshot method & The Snapshot method uses the travel time of previous bus on the same route as prediction in heavy traffic. Random Forrest, Extreme random Forrest, AdaBoost, Gradient Tree Optimisation, and combinations with snapshot methods were tested. Optimised versions use the absolute deviation error instead of mean quadratic error. &   
    RMSE, MARE, MdARE &   
    
    Snapshot method improved the accuracy MARE(\%): snapshot-optimised gradient-boost =19.06, optimised gradient-Boost =19.38 snapshot gradient-boost =19.95, gradient-boost =20.46,  extreme RF  22.05, snapshot =23.37, RF =24.11, snapshot-adaBoost =26.38, adaBoost =27.08 &  Travel time of last bus for the same segment. Headway to last bus. Day of the week. Time of day. & snapshot-optimised gradient-boost\textless optimised gradient-boost \textless Snapshot gradient-boost 
    
    \textless gradient boost \textless extreme RF \textless snapshot \textless RF \textless snapshot-adaBoost \textless adaBoost \\ \midrule
    
 \textbf{ Heghedus (2017) \cite{Heghedus2017}}   & LSTM &    
 
 NN: 1-3 hidden layers with 10 \& 20 neurons in all combinations only best reported. 
 CNN: 2 convolutional, 2 pooling, 1 fully connected, 1 dropout.  Filter 1X1. LSTM: 10 LSTM cells and two activation functions tanh and sigmoid.  & MSE, Brier score &  Values not shown. &  Arrival time, departure time, distance between stops. & LSTM \textless CNN \textless NNs of different depth\\ \midrule
    
 \textbf{ Hua et al. (2017) \cite{Hua2017}}   & NN  &     SVM with RBF kernel, \newline NN: 2 hidden layers with 10 and 5 hidden units. &   
    MAE, RMSE, MAPE &   RMSE smallest for NN, LR 10\% worse than SVM and NN. &  
    
    preceding travel time of other routes, weighted average travel time of preceding buses, total travel time including on real and virtual road. Including all features above was best.  & NN \textless SVM  \textless LR \\ \midrule
    
 \textbf{ Jeong \& Rilett (2004) \cite{Jeong2004a}}   & NN & NN: 1 Hidden layer, hidden units variable depending on route up to 15,  out=ETA & MAPE & Average improvement of models 54.2\% downtown and 48.61\% in north area (Houston)  in Comparison to LR. And 71.01\% downtown and 76.53\% north compared to LR.  &dwell time, schedule adherence, distance & NN \textless LR \\ \midrule
    
 \textbf{ Julio et al (2016) \cite{Julio2016}}   & NN & NN: 3-6-5-1,  Mixed Model: SVM to cluster the speed categories, second SVM for low speed  or NN for higher speeds were used to predict the travel speed.  & RMSE, MSE, MAPE &

 MAPE improvement: NN 7.9-44.7\% compared to algorithm which uses current speed as the speed for the next 15-30 min. Bayesian Networks  performed so poorly that results were excluded.  &  3 real time 10 min previous cell speeds, Binary if the to be predicted cell is in a corridor and 4 historical speeds for cells & NN \textless mixed model \textless SVR historical \textless current speed for next section (with exceptions) \\ \midrule
    
 \textbf{ Junyou et al (2018), \cite{Junyou2018ApplicationPrediction}}   & SVM &     Based on four days and predict the 5\textsuperscript{th}. Using RBF. &   
    Relative error. &   +/- 0.5 relative error  &  Data was collected for 4 days to predict travel time of 5\textsuperscript{th} day: traffic flow, average speed, flow density and lane occupacy. & \textless no comparison \\ \midrule
    
\textbf{Kee et al (2017), \cite{Kee2017}}   & Ensemble NN &     NN: 24-50-1,  ensemble of 10 NN, output: Binary quarter of an hour &   
    Hamming loss, Accuracy, Precision, Recall, F1 & Hamming loss: NN \textless23\% . Ensemble up to 8\% better than other methods. &  Historical arrival times, peak hour, public holiday, weekday, deployment frequency, arrival times from previous hours &  NN ensemble  \textless NN \textless DT \textless RF \textless Na\"{i}ve Bayes \\ \midrule

\textbf{Khosharavi et al (2011) \cite{Khosravi2011AIntervals}}   & NN & GA used to select numbers of neurons for each of the 2 hidden NN layers. Number of neurons between 1-10. 500 NNs were trained.  &   
    PICP, MPIL, NMPIL, CLC, R2 & NN R\textsuperscript{2}: urban 25.42-46.29, freeway: 83.73 &  Weekday travel times, and time of day in one direction of the route  for 1800 trips over 6 months.  & \textless different architectures of NNs  \\ \midrule
    
\textbf{ Kumar et al. (2017) \cite{Kumar2017}}   & KNN-Kalman &     Used KNN to identify similar trajectories and KF to predict the ETA based on the identified trajectory. &   
    MAPE, MAE &   MAPE; KNN-KF: 11.6-26.6\%, average speed model 13.49-47.58\% to 11.6-26.6\%. &  Trajectories & KNN \textless average speed \\ \midrule
    
\textbf{ Li et al. (2018) \cite{Li2018}} & SVM & 
Based on the last 30 days to reduce computational cost. SVR with radial bias function. 
& Absolute error, relative error &   Average error 30s.  &  Time period, Weather, Holiday, Position &  No comparison \\ \midrule

\textbf{ Lin \& Zeng (1999) \cite{Lin1999}}   & Historical  &     Adds average time for the next section to the time at current bus stop.  & standard least square method, maximum deviation, fluctuations & Overall deviation 2.0 &  Locations & \textless comparison to different version of algorithm  \\ \midrule

\textbf{ Lin et al (2013) \cite{Lin2013Real-TimeChina}}   & NN & NN: 11-16-1 for 4 different conditions  combined to hierarchical NN.&   
    Relative average error, Relative variance error, Relative prediction error &   
    Hierarchical NN error of 0.2min, other methods 1 min.  &  Arrival time at stop, departure time from stop, travel time between stop, headway of previous buses, time index, index of delay. &  Hiereacrhical NN (better for short distances) = NN (better for long distances) \textless Kalman (Shalaby \& Farhan 2004) \\ \midrule

\textbf{ Maiti et al (2014) \cite{Maiti2014}}   & NN & NN: 4-7-1, output:arrival at next stop &   
    Percentage error, RMSE &   Only graphs no exact values. NN has the lowest percentage error followed by historical model and SVM.  &  Bus arrival time in previous bus stop, location (latitude and longitude) of previous and target bus stops. &  NN \textless Historical \textless SVM \\ \midrule
    
\textbf{ Meng et al. (2017) \cite{Meng2017}}   & Historical & Uses current average speed and Historical values for the same section updates the historical data. & minutes &   shows actual predictions &  Real time location, average speed in section. & \textless No comparison \\ \midrule

\textbf{ Napiah \& Kamaruddin (2009) \cite{Napiah2009ArimaPrediction}}   & ARIMA &  &   
    Mean average relative error MARE, MAPPE & MAPPE 3.88-6.42 \%  &  arrival time and departure time, location of stop points, name of location , road network map, timetable information  & \textless no comparison to other methods \\ \midrule

\textbf{ Padmanaban et al. (2009) \cite{Padmanaban2009}}   & Historical & Model based on 3 days &   
    MAPE &   MAPE= 16\%. &  Travel time, Dwell time, running time for subsections & \textless no comparison \\ \midrule

\textbf{ Pan et al. (2012) \cite{Pan2012}}   & NN &     NN: 1 Hidden layer, input=10 nodes, Hidden layer= 13 units, output predicted speed. &   
    Average prediction error &   5.7\% improvement compared to historical algorithm. &  Stop location, distance, speed & NN \textless Historical data  \\ \midrule
    
\textbf{ Shalaby \& Farhan (2003) \cite{Shalaby2003}}   & KF &     KF with output travel time &   
    RMSE, Mean Relative Error, Maximum relative error & RMSE; KF =0.36 -0.109, TLRNN =0.075 -0.166, Regression =0.76 -0.220, Historical average =0.181 -0.543. &  GPS, Passengers boarding / leaving & \textless KF \textless  TLRNN \textless  Regression \textless  Historical \\ \midrule

\textbf{ Shalaby \& Farhan (2004) \cite{Shalaby2004}}   & KF &     KF with output travel time &   
    RMSE, Mean Relative Error, Maximum relative error & RMSE; KF =0.36 -0.109, TLRNN =0.075 -0.166, Regression =0.76 -0.220, Historical average =0.181 -0.543. &  GPS, Passengers boarding / leaving & \textless KF \textless  TLRNN \textless  Regression \textless  Historical \\ \midrule

\textbf{ Sinn et al. (2012) \cite{Sinn2012}}   & Kernel regression & Kernel regression with Gaussian kernel &   
    Absolute error in min & For a time horizon of 50m in absolute error <10\% &  Trajectories &  Kernel regression \textless KNN \textless Linear regression \textless Delay based \\ \midrule

\textbf{ Treethidtaphat et al. (2017),  \cite{Treethidtaphat2017} }   & DNN &     DNN: 11-7-7-7-7-1 &   
    MAE, RMSE &   MAPE of DNN 55\% lower than OLS  &  Current location, Target location, Distance, Instantaneous speed, GPS point average speed, Hour, Day & \textless DNN \textless Ordinary least square regression \\ \midrule

 \textbf{ Vanajakshi et al. (2009) \cite{Vanajakshi2009TravelBuses}}   & KF &     KF with output travel time &   
    APE &  APE: KF 9.3\%  better than 7d average &  Coordinates, speed, time & KF  \textless average \\ \midrule

\textbf{ Wang et al (2014) \cite{Wang2014}}   & RBF-NN & NN: 1 hidden exact architecture not shown &   
    MAPE &   Compared to multiple linear regression, and NN. &  Travel time, Dwell time, distance to next stop, passengers getting on/off, delay, speed to next stop. Online system live speeds, variability. & RBFNN online \textless RBFNN offline \textless  LR \\ \midrule

\textbf{ Xinghao et al (2013) \cite{Xinghao2013}}   & Exponential smoothing &     Exponential smoothing method &   
    MAE, MAPE & Algorithm based on simulated RFID ~14\% better MAPE than only GPS data. & length of link, departure time of previous stop, arrival time of previous bus stop, number of intersections on link, delay delay at intersection, distance passed when decelerating or accelerating, acceleration time from static to running speed, the deceleration time from running speed to static.  & \textless comparison of different input features \\ \midrule

\textbf{ Xu \& Ying (2017) \cite{Xu2017}}   & Clustering &     &   
    MAPE, MAE, RMSE &   Time dependent graph performed better than NN and SVM. Only graphs shown  &  Trajectories & trajectory \textless  NN \textless  SVM \\ \midrule
    
\textbf{ Yin et al. (2017) \cite{Yin2017}}   & SVM/NN & SVM:3-3-1  output=travel; NN: 3-5-1  &   
    MAE, RMSE & Both performed similarly. All predictions with MAPE around 10\%. &  weighted travel time of preceding bus of same number, weighted travel time of preceding bus of different number, average speed of objective bus &  Compared different feature numbers \\ \midrule
    
\textbf{ Yu et al. (2010) \cite{Yu2010HybridStation}}   & SVM-KF hybrid &     Use SVM to predict the baseline travel times and the KF to predict the arrival time. & RMSE &  SVM-KF hybrid performance better than actual timetable, The SVM-KF by 11.1\% better than  NN-KF &  Latest bus arrival time, with estimated baseline travel times &  SVM-KF \textless NN-KF \\ \midrule
    
\textbf{ Yu et al. (2011) \cite{Yu2011}}   & SVM &    

NN 4-5-1 SVM radial bias function. & MAE, MAPE, RMSE, R &   

R: SVM0.9, NN=0.87, kNN=0.85, LR=0.84 &  Headway to same line and last bus at stop, running time between stops of same line and last preceding bus.  &  SVM \textless NN \textless k-NN \textless LR \\ \midrule

\textbf{ Yu et al. (2017) \cite{Yu2017}}   & RFNN &     Random Forrest based on Nearest Neighbour RFNN &   
    MAE, RMSE, MAPE & MAPE for route with better performance; RFNN: 6.9\%, RF: 8.24\%, SVM: 11.16\%, KNN: 17..33\%, LR: 16.41\%  &  dwell time, running condition on the current route segment of 3 preceding buses, running condition of next segment of 3 preceding buses, for each segment, speed variance, average speed,  & RFNN \textless  SVM \textless  KNN \textless LR \\ \midrule
    
\textbf{ Zaki et al. (2013) \cite{Zaki2013}}   & Hybrid NN-KF & NN: 7-10-3-1 &   
    MSE &   NN 1.2 min MSE on route, KF 1 min on whole route. Has the MSE been confused with RMSE?. &  Day, Direction, Stations, Days Category, Weather, Avg. speed, traffic status  &  NN-KF \textless NN \\ \midrule
    
\textbf{ Zhang et al. (2015) \cite{Zhang2015}}   & Historical  &     Use the historical travel time between stops and the historical dwell time at the stops to predict overall travel time.  &   
    APE &   No comparison if further away than 5 stops error of ~60s when it gets closer it becomes smaller. Maximum error 300s.  &  Number of passengers boarding  & \textless no comparison \\

    \bottomrule
    
\end{longtable}
\end{small}
\end{centering}

Imported from Another project/table2.tex, at 12:48 pm Today